# Design of an Amplifier through Second Generation Current Conveyor

Nikhita Tripathi*[1], Nikhil Saxena *[2], Sonal Soni*[3]

*[123]PG scholars

*Department of Electronics & communication, Jaypee Institute of Information Technology Noida*

*Abstract: -* This paper describes the architecture of first and second generation current conveyor (CCI and CCII respectively) and designing an amplifier using second generation current conveyor. The designed amplifier through CCII+ can be used in various analog computation circuits and is superior in performance than the classical opamp. It provides better gain with higher accuracy.

## I. INTRODUCTION

There have been significant advances in the last decades in linear analog integrated circuit applications, since the appearance of the operational amplifier. The applications of opamp range from data converters to voltage references, analog multipliers, wave shaping circuits, oscillators and function generators. However, the classical op-amp suffers from limited gain-bandwidth product problems and from low slew rate at its output. They remain therefore unsatisfactory at higher frequencies. Current-mode circuits are used instead of voltage-mode circuits for a wide variety of applications. Current-mode does not need high voltage gain and high precision passive components, so they can be designed almost entirely with transistors. Specifically a current conveyor can provide a higher voltage gain over a larger signal bandwidth under small or large signal conditions than a corresponding op-amp circuit in effect a higher gain-bandwidth product.

Operational transconductance amplifiers (OTAs) can be used to perform the tuning but they suffer from limited output voltage swing and temperature sensitivity. Therefore, current conveyors are widely used.

From their introduction in 1968 by Smith and Sedra [1] and subsequent reformulation in 1970 current conveyors [2] have proved to be functionally flexible and versatile, rapidly gaining acceptance as both a theoretical and practical building block like filters, oscillators and amplifiers. In addition, a number of novel circuit functions and topologies have been explored. In many ways the current conveyor simplifies design in much as the same manner as the conventional operational amplifier (op-amp). Current-mode circuits have received significant attention due to their advantages compared to voltage mode circuits in terms of inherently wide bandwidth, greater linearity, wide dynamic range, simple circuitry and lower power consumption [3-4].

## II. CURRENT MODE Vs VOLTAGE MODE SIGNALS

There are several good reasons why current mode signals should be considered. Although errors can be introduced into any signal, voltage mode signals are susceptible to more than their share of problems.

- These include the impedance of the voltage source and other supply fluctuations, wire and connection resistance, the integrity of wire insulation, electrostatic and electromagnetic noise, and ground potential differences. Care must be taken not to load a voltage signal as new devices are added.

- The only real advantage of voltage mode signals is that they interface directly with D/A and A/D converters and analog multiplexing devices.

- Current mode signals are immune to loop resistance found in long wire runs and faulty connections. Additional devices generally can be added to the loop without concern for the signal, supply permitting.

- Current mode signals are relatively immune to noise, the only exception being electromagnetically induced noise which can be substantially eliminated through the proper use of shielded twisted pairs.

### III. EARLY DEVELOPMENTS

Current conveyors were first developed by Smith and Sedra(1990). The second-generation current conveyor (CCII) was developed shortly thereafter [6].

### III-A. FIRST GENERATION CURRENT-CONVEYOR (CCI)

The concept of the current-conveyor was first presented in 1968. The current-conveyor is intended as a general building block as with the operational amplifier [7]. Because of the operational amplifier concept has been current since the late 1940's, it is difficult to get any other similar concept widely accepted. However, operational amplifiers do not perform well in applications where a current output signal is needed and consequently there is an application field for current-conveyor circuits. Since current-conveyors operate without any global feedback, different high frequency behavior compared to operational amplifier circuits results.

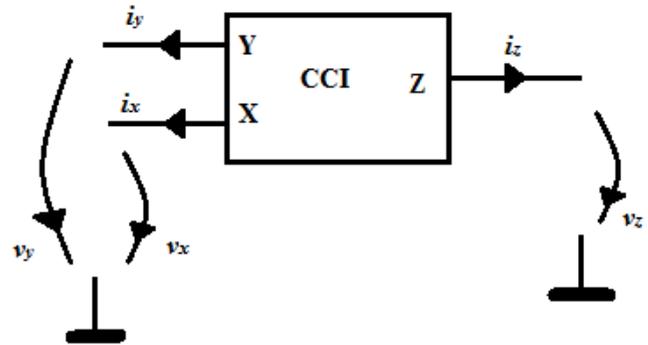

Fig. 2.1 Symbol of CCI

Current-conveyors are three-port networks with terminals X, Y and Z as represented in Figure 2.1. The network of the first generation current-conveyor CCI has been formulated in a matrix form as follows:

$$\begin{bmatrix} I_Y \\ V_X \\ I_Z \end{bmatrix} = \begin{bmatrix} 0 & 1 & 0 \\ 1 & 0 & 0 \\ 0 & 1 & 0 \end{bmatrix} \begin{bmatrix} V_Y \\ I_X \\ V_Z \end{bmatrix}$$

In other words, the first generation current conveyor CCI forces both the currents and the voltages in ports X and Y to be equal and a replica of the currents is mirrored (or conveyed) to the output port Z. The voltages at the terminals X and Y are forced to be identical. Thus the device exhibits a virtual short-circuited input characteristic at port X and a dual virtual open-circuit input characteristic at port Y.

### III-B. THE SECOND GENERATION CURRENT CONVEYOR (CCII)

The second-generation current-conveyor was developed by Sedra in 1970. A current conveyor is a building block similar to an operational amplifier and which, when used in conjunction with other components such as resistors, capacitors and diodes, can implement several useful analog sub-systems such as amplifiers, integrators, and rectifiers.

The second generation current conveyor is a three terminal device. Its symbol is shown in figure 3.1.

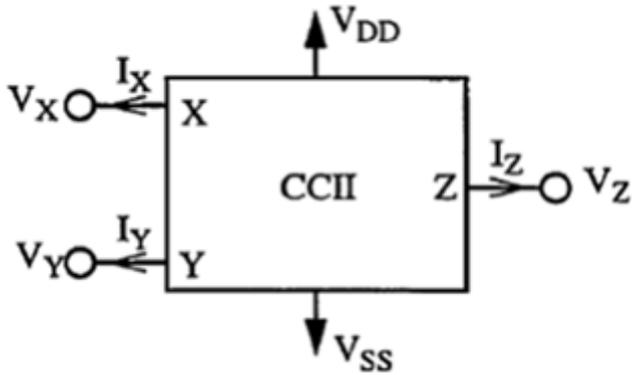

Fig 3.1 Current Conveyor Symbol and Characteristic

The current conveyor's response is given by equation

$$\begin{bmatrix} I_Y \\ V_X \\ I_Z \end{bmatrix} = \begin{bmatrix} 0 & 0 & 0 \\ 1 & 0 & 0 \\ 0 & \pm\beta & 0 \end{bmatrix} \begin{bmatrix} V_Y \\ I_X \\ V_Z \end{bmatrix}$$

Impedance is finite and must be taken into consideration in the circuit design. When a voltage is applied at node Y, that voltage is replicated at node X. This is similar to the virtual short on an op-amp. Also when a current is injected into node X, that same current gets copied into node Z. The notation CCII+ denotes a positive Z output current conveyor ($\beta = +1$) whereas CCII- denotes a negative Z output current conveyor ($\beta = -1$). Thus the terminal Y exhibits infinite input impedance. The voltage at X follows that applied to Y, thus X exhibits zero input impedance. The current supplied to X is conveyed to the high impedance output terminal Z where it is supplied with either positive polarity (in CCII+) or negative polarity (in CCII-). One model used to analyze a CCII+ is shown in figure3.2. The op-amp in unity-gain feedback configuration ensures that Vx is equal to Vy, and the current mirrors ensure that Iz is equal to $I_X$. Here the op-amp's output stage has infinite output impedance.

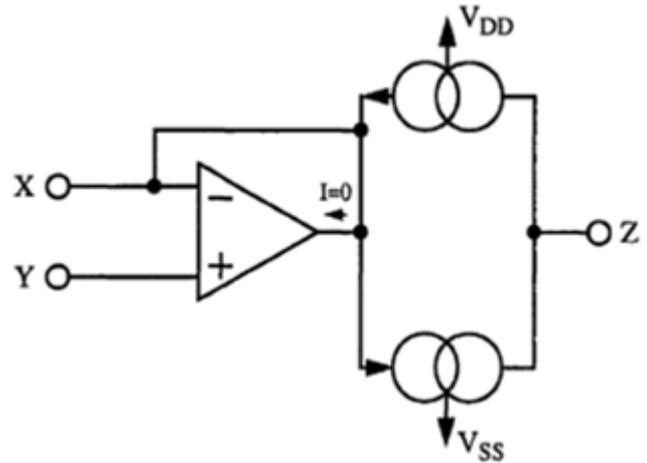

Fig 3.2 Mode1 for CCII+

A CCII- is obtained from a CCII+ by adding two current mirrors in the output stage, inverting the output current, as depicted in figure 3.4

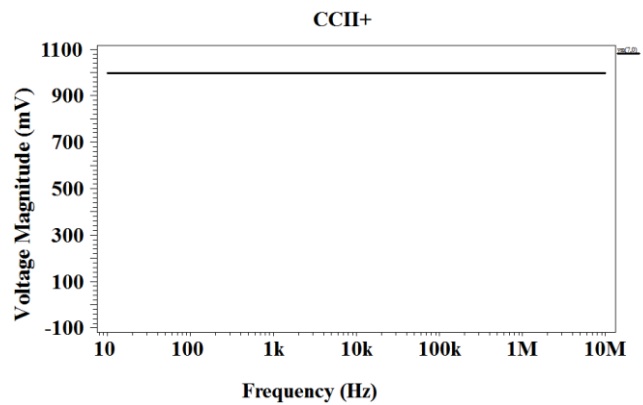

Fig.3.3 The voltage gain is 1 as seen from Frequency Analysis showing Vx=Vy of CCII

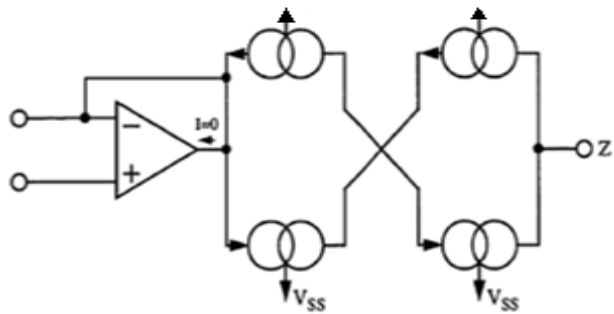

Fig 3.4 Model for CCII

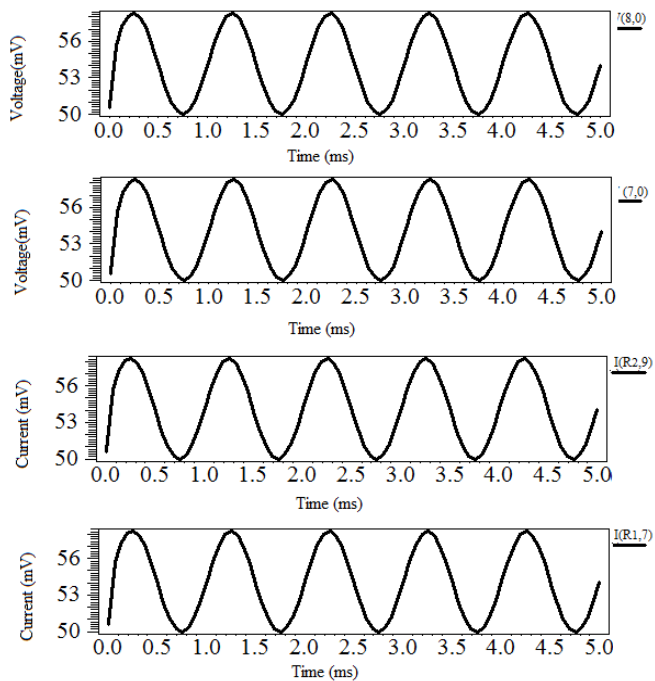

Fig. 3.5 Output waveform showing that Vx=Vy and ix=iz

IV. CMOS Implementation of CCC-II

The current controlled current conveyor (CCCII). It exhibits the same features of the CCII plus the controllability. It has been realized in different technologies, BJT, CMOS, and BICMOS [1-5]. At present, there is a growing interest in using the CCCII in various analogue signal processing applications [6-10].

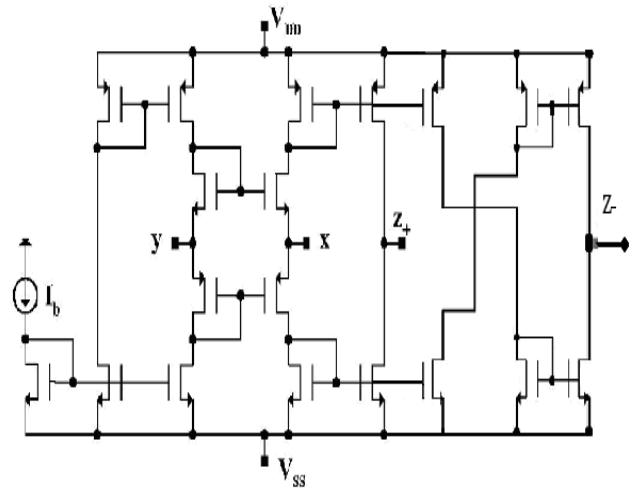

Fig 4.1 CMOS implementation of CCC-II

As a general rule, a current controlled conveyor CCCII is a translinear CCII to which the possibility of modifying the value of the DC bias current $I_b$ will confer additional properties. All the other characteristics of the CCIIs are preserved. In order to underline the advantages of the use of the controlled conveyors described here, we will compare its performance with the conventional bipolar operational transconductance amplifier (OTA) For OTA the value of gm=I/(2Vt). Comparing this value with I/R, deduced from it can be seen that for the same value of I, the transconductance of the bipolar OTA will be four times less than that of the CCCII. Thus, for the same transconductance the power consumption of the bipolar OTA will be about three time greater than with the CCCII. Second, with very high values for the collector currents of the transistors, the maximum usable frequency of the OTA will be reached sooner. This consequently indicates that the frequency performance of circuits with controlled conveyors will be much better than that for OTA implementations.

When signals are widely distributed as voltages, the parasitic capacitances are charged and discharged with the full voltage swing, which limits the speed and increases the power consumption of voltage-mode circuits. Current-mode circuits cannot avoid nodes with high voltage swing either but these are usually local nodes with less parasitic capacitances. Therefore, it is possible to reach higher speed and lower dynamic power consumption with current-mode circuit techniques.

Current-mode interconnection circuits in particular show promising performance. When the signal is conveyed as a current, the voltages in MOS-transistor circuits are proportional to the square root of the signal, if saturation region operation is assumed for the devices. Similarly, in bipolar transistor circuits the voltages are proportional to the logarithm of the signal. Therefore, a compression of voltage signal swing and a reduction of supply voltage is possible. This feature is utilized for example in log-domain filters, switched current filters, and in non-linear current-mode circuits in general.

The diagram of CCII+ is shown in figured4.2 with the nodes markad. Simulation results from SPICE tool have been shown in figure 3.3, 3.5, 4.3, 4.5. CCII+ is used as an amplifier as shown in figure 4.4. The ratio of $R_2$ and $R_1$ gives the gain of the amplifier.

$$gain, A_V = \frac{V_{OUT}}{V_{IN}} = \frac{R_2}{R_1}$$

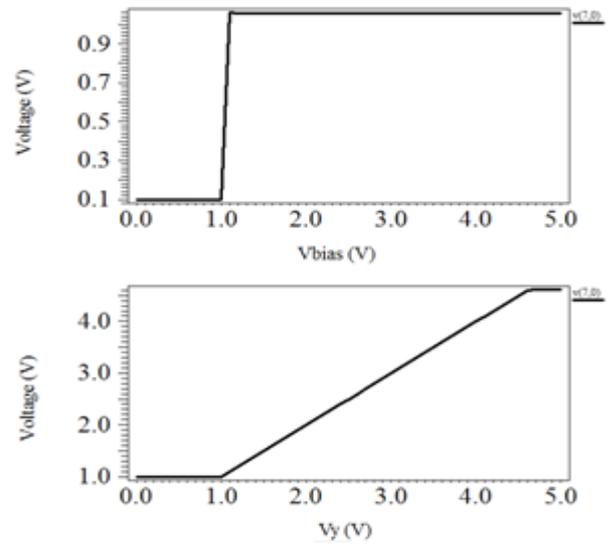

Fig. 4.3 From the transconductance curve we set the biasing point so that the transistors are in saturation.

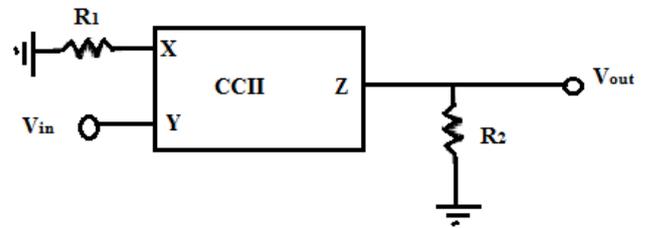

Fig 4.4 CCII as an amplifier

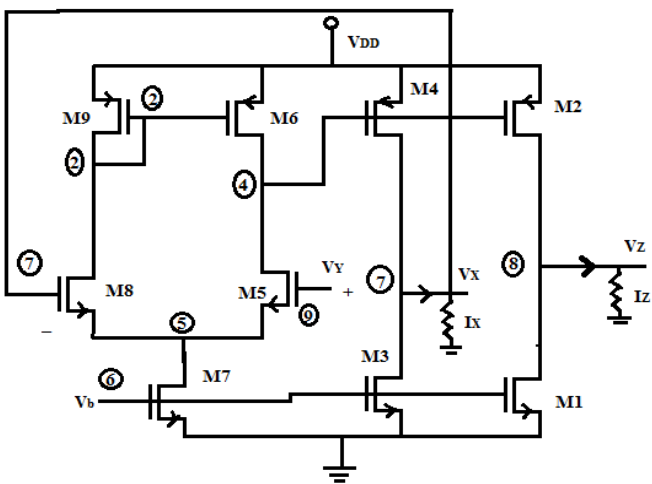

Fig 4.2   Diagram of CCII+

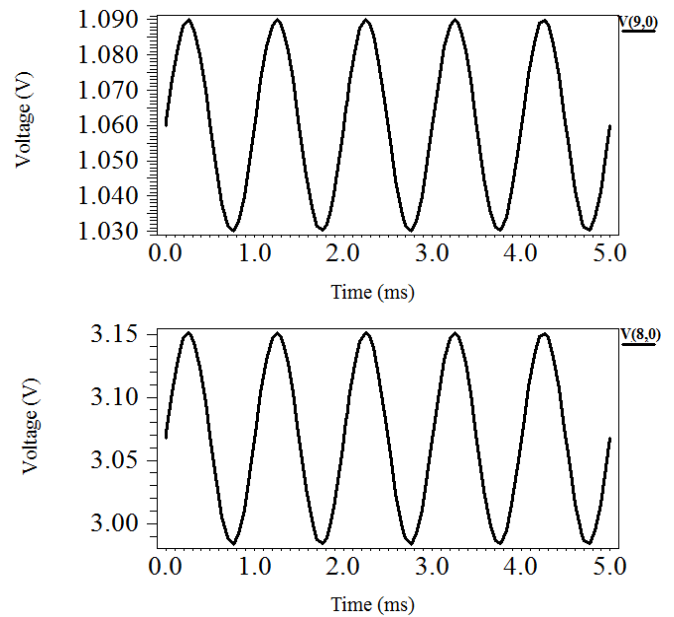

Fig.4.5 Output waveform showing amplification

V.CONCLUSION

Current-mode circuits are undoubtedly the most widely accepted operational devices in continuous time and current mode signal processing. In addition a number of novel circuit functions such as amplifier, integrator, summer, differentiator, etc. and topologies like filters and oscillator have been explored on a front of current mode analogue circuits, opening up wider area of interest.

The circuit of CCII+ is verified through TSPICE simulation results. The circuit of CCII+ and the designed amplifier is verified through TSPICE simulation results.